# Regime-Conditional Distributional Comparison of Trading Strategies:
# A GAMLSS/ZAGA Framework Applied to the S&P 500

Krzysztof Ozimek[a]

**Abstract.** Conventional comparisons of algorithmic trading strategies reduce each performance metric to a single number over the full backtest horizon, thereby discarding information about how performance varies with market conditions. This paper proposes a distributional framework that addresses this shortcoming. A walk-forward backtest of 146 out-of-sample folds on the S&P 500 (2002–2025) is used to compute the Adjusted Information Ratio ($IR^*$) for a polynomial Support Vector Machine strategy (SVMP) and a buy-and-hold benchmark (BH) in each fold. The resulting $IR^*$ sequences are modelled jointly via a Generalised Additive Model for Location, Scale and Shape (GAMLSS) with a Zero-Adjusted Gamma (ZAGA) response, with distributional parameters conditioned on market regime covariates: realised volatility and cumulative market momentum. Strategy comparison is conducted through (i) regime-specific differences in expected $IR^*$ ($\Delta E$) and its variance ($\Delta Var$), derived analytically from the fitted ZAGA parameters, and (ii) parametric bootstrap tests of three null hypotheses concerning $E(IR^*)$, $Var(IR^*)$, and their ratio, evaluated at six representative market regimes. The results demonstrate that the dominance relationship between SVMP and BH is conditional on market regime. The proposed GAMLSS/ZAGA framework constitutes a methodologically rigorous and practically interpretable alternative to conventional strategy evaluation.

# Rozkładowe porównanie strategii inwestycyjnych warunkowane reżimem rynkowym: podejście GAMLSS/ZAGA z zastosowaniem do indeksu S&P 500

**Streszczenie.** Konwencjonalne porównania algorytmicznych strategii redukują każdą miarę efektywności do pojedynczej liczby dla całego horyzontu backtesu, tym samym pomijane są informacje o tym, jak efektywność zmienia się w zależności od warunków rynkowych. W niniejszym artykule proponowane jest podejście wykorzystujące warunkowy rozkład prawdopodobieństwa eliminujące to ograniczenie. Wykorzystując testowanie wsteczne kroczące (walk-forward backtest) obejmujące 146 okien walidacyjnych (out-of-sample) na szeregu czasowym indeksu S&P 500 (2002–2025) obliczany jest w każdym oknie Skorygowany Wskaźnik Informacyjny ($IR^*$) dla strategii opartej na wielomianowej Maszynie Wektorów Nośnych (SVMP) oraz dla strategii benchmarkowej kup i trzymaj (BH). Uzyskane sekwencje $IR^*$ są modelowane łącznie za pomocą Uogólnionego Addytywnego Modelu dla Lokalizacji, Skali i Kształtu (GAMLSS) z rozkładem Zero-Adjusted Gamma (ZAGA), przy czym parametry rozkładu warunkowane są zmiennymi opisującymi reżim rynkowy: zrealizowaną zmiennością oraz skumulowanym momentum rynkowym. Porównanie strategii przeprowadzone jest poprzez (i) specyficzne dla reżimu rynkowego różnice w oczekiwanym $IR^*$ ($\Delta E$) oraz jego wariancji ($\Delta Var$), obliczone z wykorzystanie wyestymowanych parametrów ZAGA, oraz (ii) parametryczne testy bootstrap dla trzech różnych hipotez zerowych dotyczących $E(IR^*)$, $Var(IR^*)$ i ich ilorazu, przeprowadzane w sześciu reprezentatywnych reżimach rynkowych. Wyniki badania wskazują, że relacja dominacji między SVMP a BH jest uwarunkowana reżimem rynkowym. Zaproponowane podejście GAMLSS/ZAGA stanowi metodologicznie rygorystyczną i praktycznie interpretowalną alternatywę dla konwencjonalnej oceny strategii inwestycyjnych.

---

[a]Independent Researcher, Warsaw (Poland). ORCID: https://orcid.org/0009-0005-5210-9205.
E-mail: contact@drkrzysztofozimek.com

# 1. Introduction

Conventional evaluation of algorithmic trading strategies reduces the comparison to a single number per strategy – one aggregate performance metric over the entire backtest horizon – thereby collapsing the full distribution of period-by-period performance and discarding information about how performance varies with market conditions and how dispersed it is across folds. Two strategies represented by identical aggregate metrics may differ sharply in their distributional characteristics – mean, variance, and others – yet scalar point-estimate tests are blind to these differences.

The sampling properties of performance measures have been studied theoretically since Chen and Lee (1981), who showed that the Sharpe measure's distribution depends on sample size, investment horizon, and market conditions, with Kao et al. (2023) extending this to fat-tailed returns. These contributions establish formal sampling distributions for performance measures but do not model how those distributions change as a function of observable market-regime covariates.

GAMLSS – which conditions location, scale, and shape parameters simultaneously on covariates – has been applied to distributional forecasting in energy and financial markets (Serinaldi, 2011; Ugwunze et al., 2026) and extended to non-homogeneous Markov-switching regimes (Ammann et al., 2026), yet none of these contributions targets the distribution of a trading-strategy performance measure.

In the strategy-comparison literature, stochastic dominance methods – Almost First-degree Stochastic Dominance for buy-and-hold versus revised portfolios (Levy, 2025), drawdown-based dominance (Vedernikov et al., 2023), and conditional stochastic dominance (Liu & Chang, 2026) – provide rigorous pairwise rankings via CDF constraints but do not condition distributional parameters on observable market-state covariates. Machine learning strategies increasingly evaluate performance across sub-periods (Cao & Zhang, 2025), yet still summarise each regime by a scalar statistic.

This paper applies a GAMLSS with a Zero-Adjusted Gamma (ZAGA) distribution to per-fold Adjusted Information Ratio ($IR^*$) sequences from a polynomial Support Vector Machine (SVMP) strategy and a buy-and-hold (BH) benchmark. Market-regime covariates and a strategy dummy enter all three distributional parameters simultaneously, enabling inference on differences in expected $IR^*$, variance of $IR^*$, and signal-to-noise ratio of $IR^*$ across strategies – derived analytically from the estimated conditional parameters and assessed via parametric bootstrap tests. Section 2 describes data, backtest, and model; Section 3 reports results; Section 4 concludes.

## 2. Research Method

The empirical framework proceeds in three stages. First, daily S&P 500 returns are used to construct a walkforward backtest in which a polynomial Support Vector Machine (SVMP) generates trading signals that form an equity curve over the full backtest horizon. Second, a risk-adjusted performance measure – the Adjusted Information Ratio, $IR^*$ – is computed for each out-of-sample window and for a passive buy-and-hold (BH) benchmark. Third, the sequence of 146 per-fold $IR^*$ observations for each strategy is modelled within the GAMLSS framework using a Zero-Adjusted Gamma distribution, with market-regime covariates characterising each out-of-sample period. The following subsections describe each stage in detail.

### 2.1. Data

The analysis uses daily closing prices of the S&P 500 index (ticker: ^ GSPC) retrieved from Yahoo Finance via the *quantmod* package (Ryan & Ulrich, 2025) for the period 2 January 2002 to 31 December 2025. Daily log returns are computed as $r_t = \ln(P_t/P_{t-1})$. The S&P 500 is chosen as the benchmark equity index because it provides a long, liquid, and extensively studied price history that enables robust evaluation of walk-forward performance over multiple market cycles, including the dot-com recovery (2002-2004), the global financial crisis (2007-2009), the COVID-19 shock (2020), and the high-inflation period (2022-2023).

All regime covariates introduced in Section 2.3 are derived exclusively from the S&P 500 price series itself, ensuring that the model relies solely on own-series, economically interpretable measures.

## 2.2. Backtesting Procedure

The backtesting framework follows the walk-forward (rolling-window) design standard in the algorithmic trading literature (Grudniewicz & Ślepaczuk, 2025). A fixed-length in-sample (IS) window of $T_{\text{IS}} = 160$ trading days is used to train the model, followed by an out-of-sample (OOS) evaluation window of $T_{\text{OOS}} = 40$ trading days. The windows advance by a step of $\Delta =$ 40 days (non-overlapping OOS windows), yielding $B = 146$ complete folds covering the full sample period. Figure 1 illustrates the rolling-window scheme and the resulting OOS path formed by concatenating the consecutive OOS windows.

**Figure 1.** Schematic of the walk-forward (rolling window) backtesting procedure

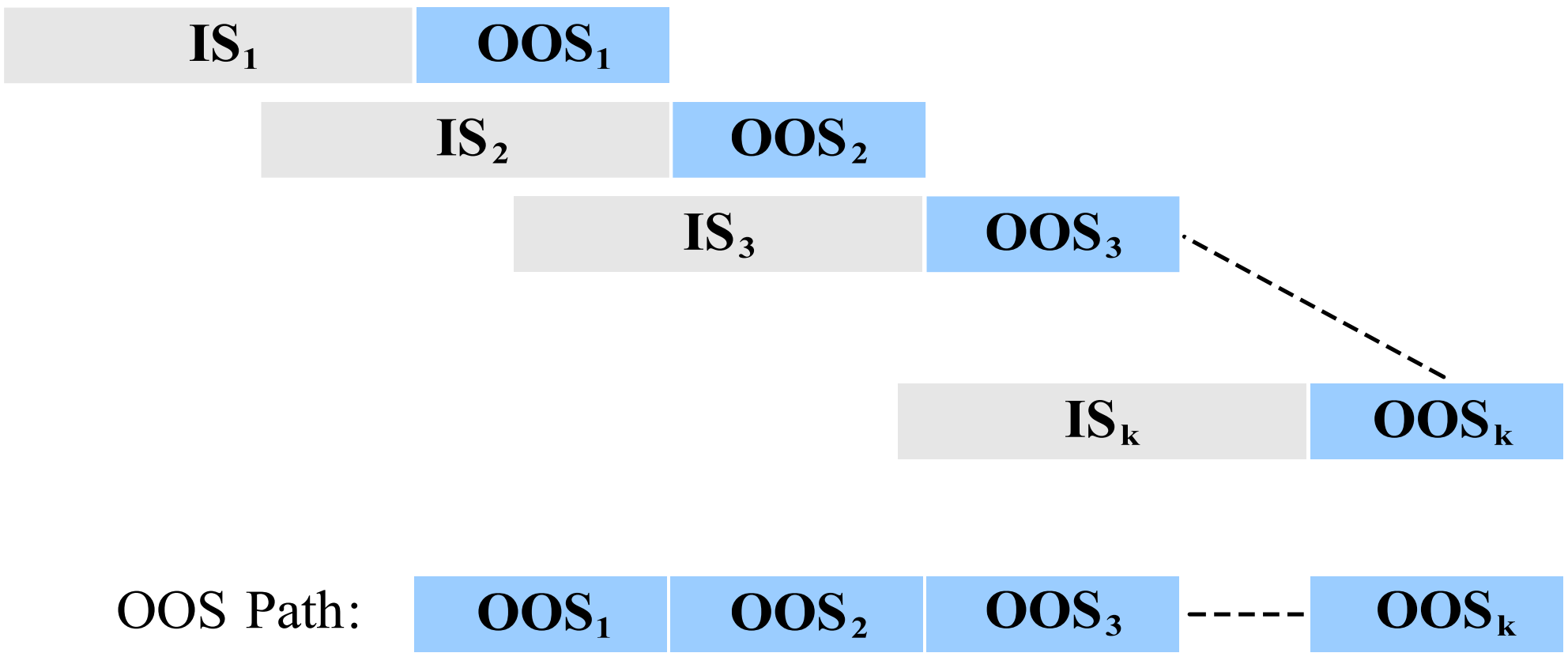

Source: adapted from Cajas (2025).

### 2.2.1. Technical Indicators

Seven technical indicators are computed from the IS price series and serve as the feature matrix $\mathbf{X} \in \mathbb{R}^{T_{\text{IS}} \times 7}$ for model training, following the selection originally proposed by Dash and Dash (2016) and adopted by Grudniewicz and Ślepaczuk (2025):

- Simple Moving Average (SMA) (Murphy, 1999) with window $n = 15$ days; the model input is the price deviation $SMA_{signal,\,t} = P_t - SMA_{15,t}$, where $\text{SMA}_{15,t} = \frac{1}{15}\sum_{i=0}^{14} P_{t-i}$;
- Moving Average Convergence/Divergence (MACD) (Appel, 2005): the MACD line is the difference between the 12-day and 26-day EMA; the model input is the MACD signal $MACD_{\text{signal},\,t} = MACD_t - signal_t$ , where $signal_t$ is the 9-day EMA of the MACD line;
- Stochastic Oscillator (Lane, 1984): fast-% K window 14, fast-%D smoothing 3, and slow-%D smoothing 3, yielding three series (fastK, fastD, slowD);
- Relative Strength Index (RSI) (Wilder, 1978) with window $n = 14$ days;
- Williams %R (WPR) (Williams, 1979) with window $n = 14$ days.

The Stochastic Oscillator contributes three series (fastK, fastD, slowD), so the total feature dimensionality is $p = 7$. All indicators are lagged by one period before use as model inputs to avoid look-ahead bias, and computed using the *TTR* package in R (Ulrich, 2023).

#### 2.2.2. Trading Model – SVMP

The predictive model is a Support Vector Machine with a polynomial kernel (SVMP), following Grudniewicz and Ślepaczuk (2025), who find SVMP achieves the best risk-adjusted performance on the S&P 500 across eight supervised learning models. The polynomial kernel used in the calculation of $f(\cdot)$ in equation (1), $k(\boldsymbol{x}_i, \boldsymbol{x}_j) = \left(c + \frac{1}{p}\boldsymbol{x}_i'\boldsymbol{x}_j\right)^d$, maps the technical indicators non-linearly, capturing their pairwise interactions without explicit feature engineering (Boser et al., 1992; Cortes & Vapnik, 1995). It is applied in regression mode to predict daily returns, with parameters set deliberately to $p = 7$ (number of input features), $c = 2$, $d = 2$. Fitting uses the *e1071* package in R (Meyer et al., 2024).

$$\hat{r}_t = f\left(SMA_{\text{signal},t}, MACD_{\text{signal},t}, fast\%K_t, fast\%D_t, slow\%D_t, RSI_t, WPR_t\right), \qquad (1)$$

where $\hat{r}_t$ is the predicted daily return at period $t$ and $f(\cdot)$ is the fitted SVMP model, so all pairwise interactions among the seven indicators enter the prediction implicitly.

#### 2.2.3. Model Inputs Transformation

Prior to being fed into the SVMP model, all technical indicator series are rescaled using a min-max normalisation to the range $[-1,1]$. The target variable (daily returns) also lies in this range in practice, so aligning the inputs makes the feature space comparable in scale and improves the numerical stability of the kernel computation. Following Han et al. (2011), the transformation for variable $x$ at period $t$ is

$$x_t' = \frac{x_t - min(x)}{max(x) - min(x)} \cdot (max_{\mathrm{norm}} - min_{\mathrm{norm}}) + min_{\mathrm{norm}} = \frac{x_t - min(x)}{max(x) - min(x)} \cdot 2 - 1, \tag{2}$$

where $x_t$ is the original value of the indicator at period $t$, $min(x)$ and $max(x)$ are the minimum and maximum values of that indicator computed over the IS window (precluding look-ahead bias), and $x_t'$ is the rescaled value passed to the model. All transformations are estimated on the IS window only and applied identically to the corresponding OOS observations, so no OOS information enters the normalisation.

#### 2.2.4. Signal Generation and Strategy Returns

At each fold, the SVMP model is fitted to the IS observations and used to produce fitted values $\hat{r}_t$ for the IS period. Three-class thresholds are derived from the empirical 40th and 60th percentiles of these fitted IS values – denoted $q_{40}$ and $q_{60}$ – and the trading signal for OOS day $t$ is defined as

$$s_t = \begin{cases} +1 & \text{if } \hat{r}_t > q_{60} \\ 0 & \text{if } q_{40} \leq \hat{r}_t \leq q_{60} \\ -1 & \text{if } \hat{r}_t < q_{40} \end{cases} \tag{3}$$

A signal of $+1$ indicates a long position, $-1$ a short position, and 0 no exposure (cash). The daily strategy return on OOS day $t$ is $r_t^{\mathrm{SIG}} = s_t \cdot r_t$, while the buy-and-hold benchmark return is simply $r_t^{\mathrm{BH}} = r_t$. The passive buy-and-hold strategy serves as the reference against which the active SVMP strategy is evaluated. Throughout the paper, BH denotes the buy-and-hold benchmark and SIG denotes the active SVMP strategy.

#### 2.2.5. Performance Measure – Adjusted Information Ratio

Strategy performance in each fold is quantified by the Adjusted Information Ratio ($IR^*$), proposed by Grudniewicz and Ślepaczuk (2025). $IR^*$ departs from the classical information ratio – defined as annualised excess return over tracking error (Bacon, 2023) – in three respects: it is benchmark-free (absolute return replaces excess return), the numerator squares the mean return and truncates at zero (penalising non-positive performance more strongly), and the denominator compounds volatility with maximum drawdown, adding an explicit capital-loss penalty absent from the classical formulation. The result is a composite measure simultaneously penalising high return dispersion and severe capital loss – two dimensions that pure mean return or Sharpe-ratio metrics also leave unaddressed. For fold $b$, the measure is defined as

$$IR_b^* = \frac{(\max\{\bar{r}_b; 0\})^2}{\hat{\sigma}_{\bar{r}_b} \cdot MDD_b}, \tag{4}$$

where $\bar{r}_b$ is the mean daily return in OOS fold $b, \hat{\sigma}_{\bar{r}_b}$ is the corresponding standard deviation, and $MDD_b$ is the maximum drawdown over the OOS window, defined as

$$MDD_b = \max_{1 \le t \le u \le T_{OOS}} \frac{V_{b,t} - V_{b,u}}{V_{b,t}}, \tag{5}$$

where $V_{b,t} = \prod_{s=1}^{t} \left(1 + r_{b,s}\right)$ is the cumulative wealth index within fold $b$, $t$ is the peak, and $u$ the subsequent trough. The floor at zero in the numerator of (4) ensures that folds with a negative mean return receive $\text{IR}^* = 0$, so the measure is strictly non-negative. All three components are computed from the same 40-day OOS window.

The non-negativity of $IR^*$ and the empirical concentration of mass at exactly zero – arising whenever $\bar{r}_b \le 0$ or when the drawdown is undefined – make the distribution of $\{IR_b^*\}_{b=1}^{B}$ structurally mixed: it has a discrete point mass at zero combined with a continuous positive tail. This motivates the GAMLSS/ZAGA modelling approach described in Section 2.3.

### 2.3. GAMLSS Framework

The statistical analysis employs the Generalised Additive Model for Location, Scale and Shape (GAMLSS), introduced by Rigby and Stasinopoulos (2005), implemented in the *gamlss* R package (Rigby & Stasinopoulos, 2005), and comprehensively described in Rigby et al. (2020). Through the *gamlss* package, GAMLSS permits wide selection of a distribution best suited to the random variable under study, and allows all distributional parameters to be modelled as functions of covariates simultaneously. Covariates thus enter the model through the distributional parameters, which in turn determine the characteristics of the random variable - such as its expected value, variance, among others. This flexibility is essential for the present application: $IR^*$ is assigned a distribution whose parameters are conditioned on regime-based covariates, so that the full shape of the $IR^*$ distribution – not only its location – can vary with market regime.

### 2.3.1. Zero-Adjusted Gamma Distribution (ZAGA)

Because $IR^*$ is non-negative and places a positive probability mass at zero, a continuous distribution on $(0, \infty)$ is insufficient. The Zero-Adjusted Gamma (ZAGA) distribution, documented in Rigby et al. (2020), provides an appropriate mixed distribution. Its probability function is (Rigby et al., 2020):

$$f_Y(y \mid \mu, \sigma, \nu) = \begin{cases} \nu & \text{if} \quad y = 0 \\ \dfrac{(1-\nu)\left(y^{\frac{1}{\sigma^2}-1} e^{-\frac{y}{\sigma^2 \mu}}\right)}{(\sigma^2\mu)^{\frac{1}{\sigma^2}} \Gamma\left(\frac{1}{\sigma^2}\right)} & \text{if} \quad y > 0 \end{cases} \tag{6}$$

for $y \geq 0$, where $\mu > 0, \sigma > 0$ and $0 < \nu < 1$.

The ZAGA family was selected in two steps. First, *fitDist()* from the *gamlss.dist* package (Stasinopoulos & Rigby, 2023) identified the Gamma distribution as best fitting the positive $(y > 0)$ observations under GAIC (Generalised AIC), motivating a zero-adjusted extension to handle the point mass at $y = 0$ present in the full $IR^*$ sequence. Second, an intercept-only model comparison of ZAGA against its natural alternative ZAIG – both zero-adjusted mixed distributions, differing only in whether the positive part follows a Gamma or an Inverse Gaussian – confirmed ZAGA as the appropriate family.

### 2.3.2. Distributional Moments

The first two moments of a ZAGA random variable $Y$ follow directly from the moment structure of zero-adjusted distributions (Rigby et al., 2020). With $\nu$ denoting the zero-probability mass:

$$E(Y) = (1-\nu)\mu, \tag{7}$$

$$Var(Y) = (1-\nu)\mu^2(\sigma^2 + \nu). \tag{8}$$

Equation (7) and (8) are the basis for the model-based estimators used in Section 2.4: the conditional expected $IR^*$ in fold $b$ is $(1-\hat{\nu}_b)\hat{\mu}_b$, and the conditional variance is $(1-\hat{\nu}_b)\hat{\mu}_b^2(\hat{\sigma}_b^2 + \hat{\nu}_b)$, both derived directly from the fitted distributional parameters.

Beyond the mean and variance, Rigby et al. (2020) provide a full characterisation of the ZAGA distribution, including expressions for the mode, skewness, excess kurtosis, moment-generating function, pdf, and cdf.

### 2.3.3. Link Functions and Linear Predictors

Three distributional parameters are modelled, each with its own link function and linear predictor. Following the GAMLSS convention (Rigby et al., 2020; Stasinopoulos & Rigby, 2017), the link functions are:

$$\log \mu_b = \eta_b^{(\mu)}, \log \sigma_b = \eta_b^{(\sigma)}, logit(\nu_b) = \eta_b^{(\nu)} \tag{9}$$

where the log link for $\mu$ and $\sigma$ enforces positivity, and the logit link for $\nu$ constrains the zero-probability to $(0,1)$. Each linear predictor $\eta_b^{(\cdot)}$ is a function of the regime covariates described in Section 2.3.4.

### 2.3.4. Regime Covariates

Two covariates characterise each out-of-sample window $b$ and serve as potential predictors of the ZAGA parameters. Both are computed exclusively from the S&P 500 daily return series $\{r_t\}_{t\in\mathcal{W}_b}$ within the OOS window $\mathcal{W}_b$:

Realised Volatility ($RV$). The within-window standard deviation of daily returns,

$$RV_b = \sqrt{\frac{1}{T_{\text{OOS}} - 1} \sum_{t\in\mathcal{W}_b} (r_t - \bar{r}_b)^2}, \tag{10}$$

where $\bar{r}_b$ is the mean OOS return in fold $b$. $RV$ is a volatility-based regime measure: high values indicate turbulent markets, low values indicate calm markets.

Momentum ($MOM$). The compounded cumulative return over the OOS window,

$$MOM_b = \prod_{t\in\mathcal{W}_b} (1 + r_t) - 1, \tag{11}$$

$MOM$ is a direction-based regime measure: positive values indicate upward-trending (bull) markets, negative values indicate downward-trending (bear) markets.

Descriptive statistics for both covariates are reported in Table 1.

### 2.3.5. Model Specifications

Dummy-variable (pooled) model. The primary specification stacks both sequences into a single dataset of $2B = 292$ observations with a binary strategy indicator $D_b = \mathbf{1}\{s = \text{SIG}\}, s \in \{\text{BH}, \text{SIG}\}$. The full candidate linear predictor for each distributional parameter takes the form

$$
\begin{aligned}
log\,\mu_b &= \beta_0^{(\mu)} + \beta_1^{(\mu)} D_b + \beta_2^{(\mu)} MOM_b + \beta_3^{(\mu)} RV_b + \beta_4^{(\mu)} (D_b \cdot MOM_b) + \beta_5^{(\mu)} (D_b \cdot RV_b) \\
log\,\sigma_b &= \beta_0^{(\sigma)} + \beta_1^{(\sigma)} D_b + \beta_2^{(\sigma)} MOM_b + \beta_3^{(\sigma)} RV_b + \beta_4^{(\sigma)} (D_b \cdot MOM_b) + \beta_5^{(\sigma)} (D_b \cdot RV_b) \quad (12) \\
logit\,\nu_b &= \beta_0^{(\nu)} + \beta_1^{(\nu)} D_b + \beta_2^{(\nu)} MOM_b + \beta_3^{(\nu)} RV_b + \beta_4^{(\nu)} (D_b \cdot MOM_b) + \beta_5^{(\nu)} (D_b \cdot RV_b)
\end{aligned}
$$

This general structure serves as the common starting point for both model selection strategies (Section 2.3.6): Strategy A and Strategy B both search over the same candidate set $\{D, MOM, RV, D \cdot MOM, D \cdot RV\}$ for each of the three equations.

#### 2.3.6. Model Selection

Variable selection within the GAMLSS framework follows the stepwise procedure of Stasinopoulos and Rigby (2017) implemented in the *gamlss* package. Two algorithms are applied:

Strategy A (*gamlss::stepGAICAll.A*) selects covariates independently for each distributional parameter, so $\mu$, $\sigma$, $\nu$ may retain different predictors.

Strategy B (*gamlss: :stepGAICAll.B*) enforces the same set of covariates across all three distributional parameters simultaneously.

The preferred model is chosen by the lowest GAIC value among all fitted models.

### 2.4. Strategy Comparison Framework

A central methodological premise of this paper is that algorithmic trading strategies should not be compared using single scalar statistics – because such summaries suppress the distributional heterogeneity that arises across market regimes. The GAMLSS/ZAGA model selected as in Section 2.3 and estimated as shown in equation (20) provides, for each fold $b$ and each strategy $s \in \{\mathrm{BH}, \mathrm{SIG}\}$, a full conditional distribution characterised by the triplet $(\hat{\mu}_{s,b}, \hat{\sigma}_{s,b}, \hat{\nu}_{s,b})$.

#### 2.4.1. Parametric GAMLSS Test

Setting $D_b = 0$ (BH) and $D_b = 1$ (SIG) in model (20) yields regime-specific parameters for each strategy. For the $\mu$- and $\nu$-equations the strategy-differentiating coefficients – those whose values differ between BH and SIG – are $\hat{\beta}_1^{(\mu)}, \hat{\beta}_4^{(\mu)}, \hat{\beta}_5^{(\mu)}$ and $\hat{\beta}_1^{(\nu)}, \hat{\beta}_3^{(\nu)}$ respectively; for the $\sigma$-equation the sole differentiating coefficient is $\hat{\beta}_1^{(\sigma)}$. The remaining coefficients determine the common level and regime sensitivity shared by both strategies. Substituting explicitly:

$$
\begin{gathered}
\hat{\mu}_{\mathrm{BH}} = exp\left(\hat{\beta}_0^{(\mu)} + \hat{\beta}_2^{(\mu)} MOM + \hat{\beta}_3^{(\mu)} RV\right), \\
\hat{\mu}_{\mathrm{SIG}} = exp\left(\hat{\beta}_0^{(\mu)} + \hat{\beta}_1^{(\mu)} + \left(\hat{\beta}_2^{(\mu)} + \hat{\beta}_4^{(\mu)}\right) MOM + \left(\hat{\beta}_3^{(\mu)} + \hat{\beta}_5^{(\mu)}\right) RV\right), \\
\hat{\sigma}_{\mathrm{BH}} = exp\left(\hat{\beta}_0^{(\sigma)}\right), \qquad \hat{\sigma}_{SIG} = exp\left(\hat{\beta}_0^{(\sigma)} + \hat{\beta}_1^{(\sigma)}\right), \\
\hat{\nu}_{\mathrm{BH}} = \frac{exp\left(\hat{\beta}_0^{(\nu)} + \hat{\beta}_2^{(\nu)} MOM\right)}{1 + exp\left(\hat{\beta}_0^{(\nu)} + \hat{\beta}_2^{(\nu)} MOM\right)}, \\
\hat{\nu}_{\mathrm{SIG}} = \frac{exp\left(\hat{\beta}_0^{(\nu)} + \hat{\beta}_1^{(\nu)} + \left(\hat{\beta}_2^{(\nu)} + \hat{\beta}_3^{(\nu)}\right) MOM\right)}{1 + exp\left(\hat{\beta}_0^{(\nu)} + \hat{\beta}_1^{(\nu)} + \left(\hat{\beta}_2^{(\nu)} + \hat{\beta}_3^{(\nu)}\right) MOM\right)}.
\end{gathered} \tag{13}
$$

The regime-specific gap in expected performance between strategies is then, from equation (7),

$$
\Delta E(MOM, RV) = E(IR^*_{\mathrm{SIG}}) - E(IR^*_{\mathrm{BH}}) = (1 - \nu_{\mathrm{SIG}})\mu_{\mathrm{SIG}} - (1 - \nu_{\mathrm{BH}})\mu_{\mathrm{BH}}, \tag{14}
$$

and the corresponding gap in variance, from equation (8),

$$
\begin{aligned}
\Delta Var(MOM, RV) &= Var(IR^*_{\mathrm{SIG}}) - Var(IR^*_{\mathrm{BH}}) \\
&= (1 - \nu_{\mathrm{SIG}})\mu^2_{\mathrm{SIG}}(\sigma^2_{\mathrm{SIG}} + \nu_{\mathrm{SIG}}) - (1 - \nu_{\mathrm{BH}})\mu^2_{\mathrm{BH}}(\sigma^2_{\mathrm{BH}} + \nu_{\mathrm{BH}}).
\end{aligned} \tag{15}
$$

Both $\Delta E$ (14) and $\Delta Var$ (15) are nonlinear functions of the regime covariates and the strategy differentiating coefficients listed above; they are evaluated numerically at specific $(MOM, RV)$ values in Section 3.

### 2.4.2. Parametric Bootstrap

While unconditional permutation tests on raw $IR^*$ values are a valid distribution-free alternative (Berry et al., 2021; Smeeton et al., 2025), they treat all folds as exchangeable, discarding the regime conditioning that is central to this paper. The parametric bootstrap, exploiting the full ZAGA structure, is therefore preferred.

Repeating over $N_b$ replications produces an empirical distribution of any summary statistic, affording the same flexibility in the choice of test statistic as classical permutation approaches (Berry et al., 2021).

Three null hypotheses are of central interest. For each replication $j = 1, \ldots, N_b, n_{\mathrm{rep}}$ synthetic $IR^*$ values are drawn from ZAGA$(\hat{\mu}_s, \hat{\sigma}_s, \hat{\nu}_s)$ evaluated at $(MOM_0, RV_0)$. Writing $\overline{IR^*}_s^{(j)}$ and $S_s^{*(j)}$ for the resulting sample mean and standard deviation of strategy $s$ in replication $j$, the test statistics are

$$T_1^{(j)} = \overline{IR^*}_{BH}^{(j)} - \overline{IR^*}_{SIG}^{(j)}, \quad (16)$$

$$T_2^{(j)} = \left(S^{*}{}_{BH}^{(j)}\right)^2 - \left(S^{*}{}_{SIG}^{(j)}\right)^2, \quad (17)$$

$$T_3^{(j)} = \frac{\overline{IR^*}_{BH}^{(j)}}{S^{*}{}_{BH}^{(j)}} - \frac{\overline{IR^*}_{SIG}^{(j)}}{S^{*}{}_{SIG}^{(j)}}, \quad (18)$$

corresponding to $\mathrm{H}_0{:}\,E(IR^*_{\mathrm{BH}}) \leq E(IR^*_{\mathrm{SIG}}), H_0{:}\,Var(IR^*_{\mathrm{BH}}) \leq Var(IR^*_{\mathrm{SIG}})$, and $\mathrm{H}_0{:}\,E(\mathrm{IR}^*_{\mathrm{BH}})/\sqrt{Var(IR^*_{\mathrm{BH}})} \leq E(IR^*_{\mathrm{SIG}})/\sqrt{Var(IR^*_{\mathrm{SIG}})}$, respectively. Each component $\overline{IR^*}_s^{(j)}/S^{*}{}_s^{(j)}$ in $T_3^{(j)}$ is the signal-to-noise ratio of $IR^*$ for strategy $s$ in replication $j$ – the mean bootstrap $IR^*$ divided by its standard deviation – so $T_3^{(j)}$ measures the difference in IR* consistency between BH and SIG. Under each $H_0$, the corresponding test statistic is non-positive. The bootstrap $p$-value is the proportion of replications in which $T_k^{(j)}$ is consistent with $H_0$ – i.e., non-positive:

$$\hat{p}_k = \frac{1}{N_b}\left\{j{:}\,T_k^{(j)} \leq 0\right\}. \quad (19)$$

A small $\hat{p}_k$ indicates that the fitted GAMLSS model rarely generates realisations in which BH meets or exceeds SIG in the relevant dimension – constituting evidence against $H_0$. No results are presented here; Section 3 reports $\Delta E$, $\Delta Var$ and bootstrap $p$-values computed in the regime-specific mode at the six quantile-matched evaluation points – Min., 1st Qu., Median, Mean, 3rd Qu., and Max. – of ($MOM$, $RV$) from Table 1. When the $\nu$-equation exhibits near-complete separation (Section 3.3), the fitted values $\hat{\nu}_{s,b}$ remain numerically stable and are used directly as inputs to *gamlss.dist::rZAGA()*.

## 3. Results

### 3.1. Estimated GAMLSS/ZAGA Model Specification

As a preliminary check, ZAGA models were estimated separately for each strategy. The BH series required regime covariates ($MOM$ and $RV$ selected for all three distributional parameters), whereas the SVMP series retained the null model under the BIC penalty – indicating that the $IR^*$ distribution of the active strategy is invariant to the market regime as characterised by $MOM$ and $RV$. This asymmetry motivates the pooled dummy-variable specification, where strategy-by-regime interaction terms directly capture differential regime sensitivity within a single likelihood.

The pooled model stacks both strategies ($n = 292$) with binary indicator $D_b = \mathbf{1}\{s = \text{SIG}\}, s \in \{\text{BH}, \text{SIG}\}$ and assumes $IR_b^* \sim \text{ZAGA}(\mu_b, \sigma_b, \nu_b)$ with link functions as in equation (9). Strategy A (independent stepwise per parameter, BIC penalty $k = \log 292$) is preferred over Strategy B by GAIC: $-1430.30$ ($df = 12$) vs. $-1421.08$ ($df = 18$), against a null dummy baseline of $-1137.74$ ($df = 6$). The estimated specification is

$$\begin{gathered} log\ \hat{\mu}_b = \hat{\beta}_0^{(\mu)} + \hat{\beta}_1^{(\mu)} D_b + \hat{\beta}_2^{(\mu)} MOM_b + \hat{\beta}_3^{(\mu)} RV_b + \hat{\beta}_4^{(\mu)} (D_b \cdot MOM_b) + \hat{\beta}_5^{(\mu)} (D_b \cdot RV_b), \\ log\ \hat{\sigma}_b = \hat{\beta}_0^{(\sigma)} + \hat{\beta}_1^{(\sigma)} D_b, \\ logit\ \hat{\nu}_b = \hat{\beta}_0^{(\nu)} + \hat{\beta}_1^{(\nu)} D_b + \hat{\beta}_2^{(\nu)} MOM_b + \hat{\beta}_3^{(\nu)} (D_b \cdot MOM_b). \end{gathered} \tag{20}$$

The three sub-equations of (20) capture qualitatively distinct dimensions of strategy difference. The $\mu$ equation reveals a regime crossover: the main effects of $MOM$ and $RV$ describe BH performance sensitivity, while the interaction terms $D \cdot MOM$ and $D \cdot RV$ measure how SVMP deviates from that baseline – with opposite signs, indicating that the two strategies respond differently to the same market regime. The $\sigma$ equation captures an unconditional dispersion difference: only $D$ enters, with no regime interaction, so the relative spread between strategies is stable across regimes. The $\nu$-equation captures near-complete separation in zero-performance probability driven by market direction: $MOM$ and $D \cdot MOM$ together imply that the probability of a zero-$IR^*$ fold differs sharply between strategies, consistent with the empirical excess of zero folds in SVMP (47.9%) relative to BH (26.7%). Estimated coefficients and model comparison statistics are reported in Table 1.

### 3.2. Data Summary

The pooled dataset comprises $n = 2B = 292$ observations (146 folds × 2 strategies). Table 1 reports the six-number summary for all four variables entering the model: the two response series ($IR_{\text{BH}}^*$ and $IR_{\text{SIG}}^*$) and the two regime covariates ($RV$ and $MOM$). The proportion of zero-$IR^*$ folds – a key feature of the ZAGA distribution – differs markedly between strategies: 26.7% for BH and 47.9% for SVMP, a gap that is directly reflected in the $\nu$-equation of model (20).

**Table 1.** Descriptive statistics of all model variables, per fold ($\boldsymbol{B = 146, n_{\text{pooled}} = 292}$)

| Statistic | $IR_{\text{BH}}^*$ | $IR_{\text{SIG}}^*$ | $RV$ | $MOM$ |
|---|---|---|---|---|
| Zero folds | 39 / 146 (26.7%) | 70 / 146 (47.9%) | – | – |
| Min. | 0.0000000 | 0.0000000 | 0.002636 | -0.250817 |

**Table 1.** (*continued*)

| 1st Qu. | 0.0000000 | 0.0000000 | 0.006341 | -0.002151 |
|---|---|---|---|---|
| Median | 0.0008801 | 0.0000036 | 0.008240 | 0.022069 |
| Mean | 0.0076762 | 0.0030820 | 0.010092 | 0.016989 |
| 3rd Qu. | 0.0092805 | 0.0021380 | 0.011673 | 0.052255 |
| Max. | 0.0773044 | 0.0551700 | 0.044738 | 0.232265 |
| Skewness | 2.5080 | 3.9623 | 3.1133 | -0.9685 |
| Excess kurtosis | 6.7995 | 21.0504 | 12.8257 | 4.1087 |

Note. $IR^*$ computed per 40-day OOS fold, 2002–2025. BH – buy-and-hold strategy; SIG – SVMP trading strategy. $RV$ = realised volatility (within-fold standard deviation of daily returns); $MOM$ = compounded 40-day return. $IR^*$ and $RV$ are non-negative by construction.
Source: author's own calculations.

### 3.3. Estimated Model Parameters

Table 2 reports the estimated coefficients of model (20) for each of the three distributional parameters. The RS (Rigby–Stasinopoulos) fitting algorithm converged in 2 cycles ($n = 292$, $df = 12$, residual $df = 280$ ); $GAIC(BIC) = -1430.30$, $global\ deviance = -1498.43$.

**Table 2.** Estimated GAMLSS/ZAGA parameters – pooled dummy-variable model, Strategy A

| Parameter / Covariate | Estimate | Std. Error | $t$ value | $Pr(>\|t\|)$ | Sig. |
|---|---|---|---|---|---|
| Panel A – $\log\mu$ equation (link: log) | | | | | |
| (*Intercept*) | -6.0478 | 0.2265 | -26.704 | < 2×10$^{-16}$ | *** |
| $D_{\mathrm{SIG}}$ | 1.7435 | 0.5183 | 3.364 | 0.0009 | *** |
| $MOM$ | 53.7625 | 3.7368 | 14.387 | < 2×10$^{-16}$ | *** |
| $RV$ | -229.0470 | 16.1053 | -14.222 | < 2×10$^{-16}$ | *** |
| $D_{\mathrm{SIG}} \cdot MOM$ | -57.4065 | 5.0024 | -11.476 | < 2×10$^{-16}$ | *** |
| $D_{\mathrm{SIG}} \cdot RV$ | 141.2516 | 47.9031 | 2.949 | 0.0035 | ** |
| Panel B – $\log\sigma$ equation (link: log) | | | | | |
| (*Intercept*) | -0.1106 | 0.0613 | -1.804 | 0.0724 | . |
| $D_{\mathrm{SIG}}$ | 0.5802 | 0.0897 | 6.466 | 4.5×10$^{-4}$ | *** |
| Panel C – $\mathrm{logit}\,\nu$ equation (link: logit) † | | | | | |
| (*Intercept*) | -4.427 | 4.385 | -1.010 | 0.3135 | |
| $D_{\mathrm{SIG}}$ | 4.369 | 4.388 | 0.996 | 0.3203 | |

**Table 2.** (*continued*)

| | | | | | |
|---|---|---|---|---|---|
| $MOM$ | -2728.595 | 2400.785 | -1.137 | 0.2567 | |
| $D_{\text{SIG}} \cdot MOM$ | 2727.187 | 2400.785 | 1.136 | 0.2569 | |

Note. Signif. codes: $^{***}p < 0.001$; $^{**}p < 0.01$; $^{*}p < 0.05$; $.p < 0.1$; (blank) $p \geq 0.1$. Global deviance: $-1498.43$ ; GAIC(BIC): $-1430.30$ $(df = 12)$; GAIC(AIC): $-1474.43$. Strategy significance: the effect of $D$ on $E(IR^*)$ is assessed by the Wald $t$-test on $\hat{\beta}_1^{(\mu)}$ (Panel A) and by the LRT (Likelihood Ratio Test) comparing model (20) against a restricted model without $D$ and its interactions. The effect of $D$ on $Var(IR^*)$ is assessed by the Wald $t$-test on $\hat{\beta}_1^{(\sigma)}$ (Panel B). † Panel C exhibits near-complete separation: $MOM$ perfectly predicts zero-$IR^*$ folds when $D_{\text{SIG}} = 1$. Coefficients and standard errors in the $\nu$-equation are numerically inflated; fitted values $\hat{\nu}_b$ remain well-defined and are used for computing $\hat{E}(IR_b^*) = (1 - \hat{\nu}_b)\hat{\mu}_b$.
Source: author's own calculations.

### 3.4. Regime-Stratified Strategy Comparison

Using the estimated coefficients of model (20), equations (14) and (15) are evaluated at six $(MOM, RV)$ pairs corresponding to the six-number summary (Min., 1st Qu., Median, Mean, 3rd Qu., Max.) reported in Table 1, yielding regime-specific point estimates of $\Delta E$ and $\Delta Var$. For each evaluation point, a parametric bootstrap ($N_b = 9{,}999$ replications, $n = 146$ draws per replication; Davison & Hinkley, 1997) is then used to compute the $p$-values $\hat{p}_1$, $\hat{p}_2, \hat{p}_3$ for the three null hypotheses defined by (16)–(19) and introduced in Section 2.4.2. Results are reported in Tables 3 and 4.

**Table 3.** Regime-specific expected $IR^*$ and its variance for BH and SVMP, with differences across six market regimes

| Regime | $E(IR^*_{\text{BH}})$ | $E(IR^*_{\text{SIG}})$ | $\Delta E$ | $Var(IR^*_{\text{BH}})$ | $Var(IR^*_{\text{SIG}})$ | $\Delta Var$ |
|---|---|---|---|---|---|---|
| Min. | 0.000000 | 0.011411 | +0.011411 | ≈ 0 | $9.551\times10^{-4}$ | $+9.551\times10^{-4}$ |
| 1st Qu. | 0.000094 | 0.004010 | +0.003915 | $7.000\times10^{-8}$ | $9.523\times10^{-5}$ | $+9.516\times10^{-5}$ |
| Median | 0.001172 | 0.003159 | +0.001986 | $1.100\times10^{-6}$ | $5.797\times10^{-5}$ | $+5.687\times10^{-5}$ |
| Mean | 0.000584 | 0.002725 | +0.002142 | $2.700\times10^{-7}$ | $4.333\times10^{-5}$ | $+4.306\times10^{-5}$ |
| 3rd Qu. | 0.002706 | 0.002136 | -0.000571 | $5.870\times10^{-6}$ | $2.589\times10^{-5}$ | $+2.002\times10^{-5}$ |
| Max. | 0.022197 | 0.000068 | -0.022130 | $3.949\times10^{-4}$ | $2.000\times10^{-8}$ | $-3.949\times10^{-4}$ |

Note. $\Delta E = E(IR^*_{SIG}) - E(IR^*_{BH})$; $\Delta Var = Var(IR^*_{SIG}) - Var(IR^*_{BH})$. Regime points are quantile-matched $(MOM, RV)$ pairs from Table 1. BH – buy-and-hold strategy; SIG – SVMP trading strategy. $\approx 0$ – $Var(IR^*_{BH})$ falls below $10^{-8}$; near-complete separation at the Min. regime drives all BH bootstrap draws to zero.
Source: author's own calculations.

Table 3 reveals a momentum-conditioned performance reversal. At low and negative momentum (Min. through Mean), $\Delta E > 0$ : SVMP generates higher expected $IR^*$ than BH, with the largest advantage at Min. ($\Delta E = +0.011$), where $\hat{v}_{\text{BH}} \to 1$ under near-complete separation – the model implies BH produces a near-zero expected $IR^*$ in almost every fold under strong negative momentum, while SVMP continues to generate positive expected performance. At high positive momentum (3rd Qu. and Max.), the sign reverses: $\Delta E < 0$, with BH dominating massively at Max. ($\Delta E = -0.022$), where a trending market rewards a passive strategy over an ML-based signal. Regarding variance, $\Delta Var > 0$ at all regimes except Max., meaning SVMP carries structurally higher distributional spread than BH across most of the momentum distribution – the exception at Max. reflects the large BH variance arising from the strong positive-return episodes that drive the maximum-momentum regime.

Table 4 complements the analytical results of Table 3 with distributional evidence from the parametric bootstrap. For $\hat{p}_1$, testing $\text{H}_0^{(1)}: E(IR^*_{\text{BH}}) \leq E(IR^*_{\text{SIG}})$, the bootstrap confirms SVMP dominance at low-tomid momentum: $\hat{p}_1 = 1.000$ from Min. through Mean, meaning not a single replication out of 9,999 produced a mean BH draw exceeding the mean SVMP draw – $\text{H}_0^{(1)}$ is strongly supported and cannot be rejected. The picture reverses at Max. ( $\hat{p}_1 = 0.000$ ), where $\text{H}_0^{(1)}$ is rejected and BH dominates completely in expected $IR^*$. At the 3rd Qu., $\hat{p}_1 = 0.114$ signals a transitional regime in which BH begins to pull ahead but the evidence against $\text{H}_0^{(1)}$ is not yet conclusive.

**Table 4.** Bootstrap p-values for three null hypotheses on strategy performance, across six market regimes

| Regime | $\hat{p}_1$ | $\hat{p}_2$ | $\hat{p}_3$ | $T_3$ valid (%) |
|---|---|---|---|---|
| Min. | 1.0000 | 1.0000 | NA | 0.0 |
| 1st Qu. | 1.0000 | 1.0000 | 0.8768 | 100.0 |
| Median | 1.0000 | 1.0000 | 0.0000 | 100.0 |
| Mean | 1.0000 | 1.0000 | 0.0000 | 100.0 |
| 3rd Qu. | 0.1143 | 0.9979 | 0.0000 | 100.0 |
| Max. | 0.0000 | 0.0000 | 0.0000 | 100.0 |

Note. Regime points are quantile-matched ($MOM$, $RV$) pairs from Table 1. BH – buy-and-hold; SIG – SVMP trading strategy. NA in $\hat{p}_3$ indicates that the signal-to-noise ratio test statistic $T_3$ (eq. 18) is undefined at this regime because $\hat{v}_{\text{BH}} \to 1$ under near-complete separation, causing all BH bootstrap draws to equal zero and $\text{sd}(IR^*_{\text{BH}}) = 0$ in every replication. $T_3$ valid (%): percentage of the $N_b = 9{,}999$ replications in which $sd(IR^*) > 0$ for both strategies simultaneously; $\hat{p}_3$ is computed over these replications only.
Source: author's own calculations.

For $\hat{p}_2$, testing $\mathrm{H}_0^{(2)}: Var(IR^*_{\mathrm{BH}}) \leq Var(IR^*_{\mathrm{SIG}})$, SVMP carries higher variance at all regimes except Max. ( $\hat{p}_2 = 0.000$ ), where $\mathrm{H}_0^{(2)}$ is rejected – BH variance is inflated by large positive-return episodes at peak momentum, exceeding that of SVMP. For $\hat{p}_3$, testing $\mathrm{H}_0^{(3)}: E(IR^*_{\mathrm{BH}})/\sqrt{Var(IR^*_{\mathrm{BH}})} \leq E(IR^*_{\mathrm{SIG}})/\sqrt{Var(IR^*_{\mathrm{SIG}})}$, the results are striking: despite SVMP's mean advantage at low momentum, $\mathrm{H}_0^{(3)}$ is rejected from Median onward ( $\hat{p}_3 = 0.000$ ), meaning BH achieves a superior risk-adjusted ratio – SVMP's variance penalty swamps its return advantage. At 1st Qu., $\hat{p}_3 = 0.877$ means $\mathrm{H}_0^{(3)}$cannot be rejected: in very low momentum both strategies generate near-zero returns, making the signal-to-noise ratio comparison inconclusive. At Min., $\hat{p}_3$ is undefined: near-complete separation forces $\hat{\nu}_{\mathrm{BH}} \to 1$, producing all-zero BH draws in every replication and making $sd(IR^*_{\mathrm{BH}}) = 0$ ($T_3\ valid\ = 0\%$); the analytical $\Delta E = +0.011$ from Table 3 remains the relevant comparison at this extreme.

## 4. Conclusions

This paper proposes a distributional framework for comparing algorithmic trading strategies, replacing single aggregate performance metrics with regime-conditioned full distributions. A GAMLSS/ZAGA model fitted to fold-level $IR^*$ sequences of SVMP and BH – with distributional parameters conditioned on realised volatility and momentum – reveals that the dominance relationship between strategies is strictly regime-dependent: SVMP outperforms BH in expected $IR^*$ under low and negative momentum (with decisive bootstrap support from Median through Min.), while BH dominates under strong positive momentum. SVMP exhibits higher variance in $IR^*$ across most regimes. Analytical regime-specific moments and the parametric bootstrap jointly confirm these patterns, with null hypotheses strongly retained in some regimes and decisively rejected in others. They provide a statistically evidenced, more informative basis for strategy evaluation than scalar metrics.

The scope of the conclusions is constrained by several limitations. The analysis covers a single equity index over one historical period, a single ML model class, and a performance measure gross of transaction costs. Bootstrap inference depends on the ZAGA distributional assumption, and near-complete separation at the most adverse regime introduces numerical instability in the signal-to-noise test statistic, requiring a validity-filtering step.

Natural extensions include replacing linear regime predictors with smooth additive terms and broadening the covariate set beyond realised volatility and momentum to macro-financial indicators, market sentiment indices, or latent states from hidden Markov models; where strategies account for trading frictions, transaction costs and bid-ask spreads could themselves enter the GAMLSS equations as covariates. The dummy-variable structure generalises to multi-strategy and multi-asset settings. Applied prospectively with rolling estimation and real-time regime classification, the framework would support adaptive strategy selection conditional on current market state.

Two further directions concern interpretability and simulation-based risk analysis. When covariates enter as linear or transformed terms, GAMLSS provides intrinsic explainability: each covariate's contribution to each distributional parameter is directly quantifiable from the estimated coefficients, positioning the framework as an explainable-AI (XAI) tool for strategy evaluation. When smoothers or neural network components are used instead, this direct readability is lost – but simulation remains intact. In both cases, extreme or adverse covariate values can be fed through the fitted model and synthetic performance metric samples drawn from the resulting conditional distributions, enabling scenario analysis and stress tests of trading strategy robustness. The connection between distributional regression, XAI, and simulation-based risk analysis warrants explicit development in future work.

## Disclaimer

The author used Claude (Anthropic), through the Cowork interface, to assist in writing the R scripts used for data acquisition, data cleaning, and statistical calculations, and to assist in editing and polishing the text of the article. The author takes full responsibility for the content of the article, including the parts of the analysis developed with the assistance of this tool, and bears sole responsibility for any breach of publication-ethics standards.